# Complete Hamiltonian description of wave-like features in classical and quantum physics

A. Orefice*, R. Giovanelli and D. Ditto

Università di Milano - Di.Pro.Ve. - *Via Celoria, 2, 20133 - Milano (Italy)*

**Abstract** - The analysis of the Helmholtz equation is shown to lead to an exact Hamiltonian system of equations describing in terms of ray trajectories a very wide family of wave-like phenomena (including diffraction and interference) going much beyond the limits of the standard geometrical optics approximation, which is contained as a simple limiting case.
Due to the fact that the time independent Schrödinger equation is itself a Helmholtz-like equation, the same mathematical solutions holding for a classical optical beam turn out to apply to a quantum particle beam, leading to a complete system of Hamiltonian equations which provide a set of particle trajectories and motion laws containing as a limiting case the ones encountered in classical Mechanics.

## 1. Introduction

The present paper aims to propose a simple Hamiltonian approach holding, in principle, for a wide family of wave-like phenomena, including both classical and quantum features.
Its main elements of novelty are the following:

### 1.a) *A step beyond the geometrical optics approximation*

It is often believed that the "naïve" concept of optical *rays* applies only to a very limited set of physical cases which may be ascribed to the so-called *geometrical optics approximation*, while more general and complex phenomena (such as diffraction and interference) would necessarily require a fully wave-like treatment. In the **first** (classical) part of our work (Sects.2 and 3) we show that this commonplace is not correct. Since many important optical phenomena (such as interference or diffraction patterns) can be most conveniently analyzed in time-independent conditions, we start here from the **Helmholtz** equation, from which we obtain, *without any omissions or approximation*, a Hamiltonian ray-tracing set of equations providing the exact description in term of **rays** (including both their *geometry* and their *motion law*) of a family of wave phenomena much wider than that allowed by the standard geometrical optics, which is contained as a simple limiting case. We stress, in particular, the strong dependence of the behaviour of a ray beam on its launching conditions.

### 1.b) *A "new" property of optical rays*

The rays of an optical beam are shown to be mutually correlated (in a kind of self-refractive behaviour, strongly dependent on the transverse intensity distribution of the beam) by a term acting perpendicularly to the rays themselves, and determining therefore their geometry without altering the amplitude of their velocity. This property - which is discovered here for the *first time* - is shown to provide, moreover, a basic tool for the numerical solution of the Hamiltonian ray-tracing system.

### 1.c) *Geometrical coincidence between classical optical rays and quantum particle trajectories*

And here begins the **second** (quantum) part (Sects.**4-8**) of our paper. We recall in fact that the particle trajectories encountered in Classical Mechanics were often suggested to constitute the *geometrical optics approximation* of the wave-like behaviour of the particles themselves. This suggestion was aimed, of course, to enforce the necessity of abandoning, in general, the very concept of particle trajectories, just as the ray picture was supposed to collapse in standard optics. The basic consequence of this philosophy is *quantum indeterminism*, according to which no position *and* momentum can be exactly and simultaneously ascribed to a particle.
Since, however - thanks to our previous extension of the concept of "ray" much beyond the limits of the geometrical optics approximation - no collapse of the optical ray description does occur, and since, after all, the time-independent **Schrödinger** equation is itself a **Helmholtz**-like equation, it is quite natural to expect, now, that the same peculiar properties discovered here in Classical Optics may be extended to Quantum Mechanics.
This expectation, indeed, is immediately satisfied, and the trajectories and motion laws of a quantum particle beam turn out to be provided by a dynamical Hamiltonian system *mathematically coinciding* (in suitable dimensionless variables) with the system found in the previous optical case, and involving therefore - in correspondence with the same boundary conditions - the same geometrical trajectories, with the same transverse correlation property discovered in the classical case.

### 1.d) *Complete Hamiltonian quantum description*

This Hamiltonian system provides a complete description of the **quantum** motion of a particle beam, showing no trace of probabilistic features and containing the laws of **classical** Dynamics as a limiting case (just as the geometrical optics approximation turned out to be a particular case of our exact ray approach).

Although, of course, neither the **Helmholtz** equation nor the time-independent **Schrödinger** equation can directly describe any propagation phenomenon, they provide (for any assigned medium, and for any set of boundary conditions) a fixed trajectory frame (i.e. a fixed "weft" of "rails"), which is determined at the very outset in a way reminding Fermat and Maupertuis variational principles, and along which each particle (or ray) moves according to well-defined motion laws. The trajectory frame, in its turn, does not depend (for not-interacting particles or rays), on the intensity of the beam, which could even be reduced to a single particle at a time: a peculiar feature which has often induced to speak of single particle self-diffraction.

We present here a numerical solution for the diffraction of a beam of rays and/or particles trajectories through a single slit, stressing its strong dependence on the launching conditions of the beam itself.

### 1.e) *Origin of the so-called "quantum potential"*

Within the mathematical coincidence observed between classical rays and quantum trajectories, an important feature is represented by the fact that the same mathematical term omitted in the (classical) geometrical optics approximation and taken into account here in its exact version, turns out to give rise, in the quantum case, to the so-called "quantum potential" of the **de Broglie** and **Bohm** theory [1-5]. Such a term is not, therefore, due to a basically quantum feature, but to the structure itself of Helmholtz-like equations, present in classical as well as in quantum waves. Our choice, therefore, of a *time-independent* approach contributes to shed a new light on quantum features.

### 1.f) *Comparison with other deterministic approaches*

To our knowledge, any other "deterministic" approach to the problem of quantum dynamics (such as that of **de Broglie** and **Bohm** [1-5], together with its more recent developments [6-11]) starts from the *time-dependent* Schrödinger equation, sometimes including fluid-like and/or probabilistic considerations. This generally leads to equations whose solution is a hard - and often obscure - task, both from a logical and from a numerical point of view, even though there exists at least one line of thought which appears to minimize difficulties [10, 11].

The basic aim of our *time-independent* approach, on the other hand, is to arrive at complete set of dynamical laws, for a *limited but significant set of cases*, in a simple (but not simplistic) way, avoiding the most general treatments and the most complex phenomenology, which would only hinder any solution attempt, without providing - in such topics as interference and diffraction - any substantial advantage.

## 2. Helmholtz equation and geometrical optics

In order to establish our mathematical formalism, let us start from a classical case of wave-like behaviour.

Although many kinds of physical waves would lend themselves to the considerations we have in mind here, we shall refer, in order to fix ideas, to a monochromatic electromagnetic wave beam, with a time dependence $\div \exp(i\omega t)$, travelling through an isotropic and inhomogeneous dielectric medium. Its basic features are accounted for by the Helmholtz equation

$$\nabla^2 \psi + (n k_0)^2 \psi = 0 \ , \qquad (1)$$

where $\qquad \nabla^2 = \dfrac{\partial^2}{\partial x^2} + \dfrac{\partial^2}{\partial y^2} + \dfrac{\partial^2}{\partial z^2} \ ;$

$\psi$ represents any component of the electric or magnetic field; $n(x,y,z)$ is the refractive index of the medium, and

$$k_0 \equiv \frac{2\pi}{\lambda_0} = \frac{\omega}{c} \ , \qquad (2)$$

with obvious meaning of $\lambda_0$ and $c$. The phase velocity is given, in its turn, by

$$v_{ph}(x,y,z) = c/n(x,y,z) . \qquad (3)$$

Because of its time-independence, eq.(1) doesn't directly describe, of course, any propagation phenomenon: it only determines, together with the boundary conditions, the fixed space frame where propagation occurs.

By performing the quite general replacement

$$\psi(x,y,z) = R(x,y,z) e^{i\varphi(x,y,z)} \ , \qquad (4)$$

with real $R(x,y,z)$ and $\varphi(x,y,z)$, and separating the real from the imaginary part, eq.(1) splits into the well known [12] and strictly equivalent system of coupled equations

$$\begin{cases} (\underline{\nabla}\varphi)^2 - (nk_0)^2 = \dfrac{\nabla^2 R}{R} \\ \underline{\nabla} \cdot (R^2 \underline{\nabla}\varphi) = 0 \end{cases} \qquad (5)$$

where $\underline{\nabla} \equiv \partial/\partial \boldsymbol{r} \equiv (\partial/\partial x, \partial/\partial y, \partial/\partial z)$, and the second of eqs.(5) expresses the constancy of the flux of the vector $R^2 \underline{\nabla}\varphi$ along any tube formed by the lines of $\underline{\nabla}\varphi$ itself.

When the space variation length, $L$, of the amplitude $R(x,y,z)$ may be assumed to satisfy the condition $k_0 L >> 1$, the first of eqs.(5) is well approximated by the *eikonal equation*

$$(\underline{\nabla}\varphi)^2 \cong (nk_0)^2 \ , \qquad (6)$$

decoupled from the second of eqs.(5) (whose presence is generally neglected) and allowing the so-called *geometrical optics approximation*, which describes the wave propagation in terms of "rays" travelling along the field lines of the *wave vector*

$$\boldsymbol{k} = \underline{\nabla}\varphi \qquad (7)$$

*independently from the amplitude distribution R(x,y,z) of the beam.* To be sure, by multiplying eq.(6), for convenience, by the constant factor $\frac{c}{2k_0}$, we obtain the relation

$$D(\mathbf{r},\mathbf{k}) \equiv \frac{c}{2k_0}[k^2 - (nk_0)^2] \cong 0, \qquad (8)$$

(where $\mathbf{r} \equiv (x,y,z)$), whose differentiation

$$\frac{\partial D}{\partial \mathbf{r}} \cdot d\mathbf{r} + \frac{\partial D}{\partial \mathbf{k}} \cdot d\mathbf{k} = 0 \qquad (9)$$

directly provides, for any assigned refractive function $n(\mathbf{r})$, both the geometrical form of the rays and their motion law in the simple Hamiltonian form

$$\begin{cases} \dfrac{d\mathbf{r}}{dt} = \dfrac{\partial D}{\partial \mathbf{k}} = \dfrac{c\mathbf{k}}{k_0} \\ \dfrac{d\mathbf{k}}{dt} = -\dfrac{\partial D}{\partial \mathbf{r}} = \dfrac{c}{2k_0}\dfrac{\partial}{\partial \mathbf{r}}(nk_0)^2 \end{cases} \qquad (10)$$

where a ray velocity $\mathbf{v}_{ray} = \dfrac{c\mathbf{k}}{k_0}$ is implicitly defined. We may observe that $v_{ray} \equiv |\mathbf{v}_{ray}| = c$ when $k = k_0$, and that $v_{ray} v_{ph} = c^2$.

We conclude the present Section by recalling **Fermat**'s variational principle, according to which any optical ray travelling between two points *A, B* shall follow a trajectory satisfying the condition

$$\delta \int_A^B k\, ds = 0, \qquad (11)$$

where $k = |\mathbf{k}|$ and *ds* is an element of a (virtual) line connecting *A* and *B*.

### 3. Beyond the geometrical optics approximation

Let us consider now the first of eqs.(5) in its *complete* form, arriving therefore at the *exact* relation, generalizing the function $D(\mathbf{r},\mathbf{k})$ of eq.(8),

$$D(\mathbf{r},\mathbf{k}) \equiv \frac{c}{2k_0}[k^2 - (nk_0)^2 - \frac{\nabla^2 R}{R}] = 0, \qquad (12)$$

whose differentiation, formally coinciding with eq. (9), leads to the *exact* Hamiltonian ray-tracing system

$$\begin{cases} \dfrac{d\mathbf{r}}{dt} = \dfrac{\partial D}{\partial \mathbf{k}} = \dfrac{c\mathbf{k}}{k_0} \\ \dfrac{d\mathbf{k}}{dt} = -\dfrac{\partial D}{\partial \mathbf{r}} = \dfrac{c}{2k_0}\dfrac{\partial}{\partial \mathbf{r}}[(nk_0)^2 + \dfrac{\nabla^2 R}{R}] \end{cases} \qquad (13)$$

The system (13) completely avoids the standard approximation of geometrical optics, although fully retaining the idea of electromagnetic "rays" travelling along the field lines of $\mathbf{k} \equiv \underline{\nabla}\varphi$, which depend, however, on the wave amplitude distribution *R(x,y,z)* of the beam. In order to exploit this dependence we must recall the presence of the *second* of eqs. (5), which may be written in the form

$$\underline{\nabla}\cdot(R^2\underline{\nabla}\varphi) \equiv 2R\,\underline{\nabla}R\cdot\underline{\nabla}\varphi + R^2\underline{\nabla}\cdot\underline{\nabla}\varphi = 0 \qquad (14)$$

Since no new ray trajectory may suddenly arise in the space region spanned by the beam, we must have, of course, $\underline{\nabla}\cdot\underline{\nabla}\varphi = 0$, so that eq.(14) splits into the system

$$\begin{cases} \underline{\nabla}\cdot\underline{\nabla}\varphi = 0 \\ \underline{\nabla}R\cdot\underline{\nabla}\varphi = 0 \end{cases} \qquad (15)$$

where the second equation is automatically entailed by the first one. The values of the function *R(x,y,z)* are therefore constant (i.e. "transported") along the field lines of $\mathbf{k} \equiv \underline{\nabla}\varphi$, to which $\underline{\nabla}R$ turns out to be *perpendicular*, and this transverse character is shared by the gradient $\dfrac{\partial}{\partial \mathbf{r}}\dfrac{\nabla^2 R}{R}$. The amplitude $v_{ray}$ of the ray velocity shall remain, in vacuum, equal to *c* all along its trajectory, because such a gradient may only modify the *direction*, but not the *absolute value*, of the wave vector $\mathbf{k}$: the only possible changes of $|\mathbf{k}|$ could be due, in a medium different from vacuum, to its refractive function *n(x,y,z)*.

Thanks to its constancy along each ray trajectory the function *R(x,y,z)*, once assigned on the launching surface from where the ray beam is assumed to start, may be numerically built up step by step, together with its derivatives, in the whole region crossed by the beam. As we shall see in Sect.**7**, indeed, the *exact* equation system (13) lends itself to a practicable numerical solution, even in physical cases where the standard geometrical optics *approximation* is completely inapplicable.

### 4. The time-independent Schrödinger equation

The *classical* motion of a mono-energetic beam of non-interacting particles of mass *m* through a force field deriving from a potential energy *V(x,y,z)* not explicitly depending on time may be described for each particle of the beam, as is well known, by means of the so-called "reduced" (or "time-independent") Hamilton-Jacobi equation [**10**]

$$(\underline{\nabla}S)^2 = 2m(E - V), \qquad (16)$$

where *E* is the total energy, and one of the main properties of the function *S(x,y,z)* is that the particle momentum is given by

$$\mathbf{p} = \underline{\nabla}S. \qquad (17)$$

Recalling **Maupertuis**' variational principle

$$\delta \int_A^B p\, ds \equiv 0, \qquad (18)$$

with $p = |\mathbf{p}|$, the formal analogy between eqs.(6,7,11) on one side, and eqs.(16-18) on the other side, suggests, as is well known, that the *classical* particle trajectories could constitute the *geometrical optics approximation* of an equation (analogous to the Helmholtz eq.(1)), which is immediately obtained by means of the substitutions

$$\begin{cases} \varphi = \dfrac{S}{a} \quad \text{and therefore} \\ \boldsymbol{k} = \underline{\nabla}\varphi = \dfrac{\underline{\nabla}S}{a} = \dfrac{\boldsymbol{p}}{a}; \\ k_0 \equiv \dfrac{2\pi}{\lambda_0} = \dfrac{\sqrt{2mE}}{a} \equiv \dfrac{p_0}{a} \\ n^2(x,y,z) = 1 - \dfrac{V(x,y,z)}{E} \end{cases} \quad (19)$$

where the parameter *a* represents a constant *action* whose value is a *priori* arbitrary - as far as the relations (19) are concerned - but is imposed by the history itself of Quantum Mechanics :

$$a = \hbar \cong 1.0546 \times 10^{-27} erg \cdot s . \quad (20)$$

The equation obtained from the Helmholtz equation (1) by means of the substitutions (19) and (20) takes up the form

$$\nabla^2 \psi + \dfrac{2m}{\hbar^2}(E-V)\psi = 0, \quad (21)$$

which is the standard *time-independent* Schrödinger equation. By applying now to eq.(21) the same procedure leading from eq.(1) to eqs.(5), and assuming therefore

$$\psi(x,y,z) = R(x,y,z)e^{iS(x,y,z)/\hbar} \quad (22)$$

eq.(21) splits [**13**] into the coupled system

$$\begin{cases} (\underline{\nabla}S)^2 - 2m(E-V) = \hbar^2 \dfrac{\nabla^2 R}{R} \\ \underline{\nabla}\cdot(R^2 \underline{\nabla}S) = 0 \end{cases} \quad (23)$$

By taking the gradient of the first of eqs.(23) we get moreover

$$(\dfrac{\underline{\nabla}S}{m}\cdot\underline{\nabla})\dfrac{\underline{\nabla}S}{m} + \dfrac{\underline{\nabla}V}{m} = \dfrac{\hbar^2}{2m^2}\underline{\nabla}(\dfrac{\nabla^2 R}{R}). \quad (24)$$

Eq.(24), together with the second of eqs.(23), is often interpreted as describing, in the "classical limit" $\hbar \to 0$ (whatever such a limit may mean), a "fluid" of particles with mass *m* and velocity $\dfrac{\underline{\nabla}S}{m}$, moving in an external potential *V(x,y,z)*: an interpretation consistent with the probabilistic character usually ascribed to the Schrödinger equation.

**5. Hamiltonian description of quantum particle motion**

Let us now observe that, by simply maintaining eq.(17), the first of eqs.(23) may be written in the form of a generalized, time-independent Hamiltonian

$$H(\boldsymbol{r},\boldsymbol{p}) \equiv \dfrac{p^2}{2m} + V - \dfrac{\hbar^2}{2m}\dfrac{\nabla^2 R}{R} = E, \quad (25)$$

including the "new" and crucial term $-\dfrac{\hbar^2}{2m}\dfrac{\nabla^2 R}{R}$, coinciding with the so-called *quantum potential* of the **de Broglie-Bohm** theory [**1-5**].
By differentiating eq.(25) we get the relation

$$\dfrac{\partial H}{\partial \boldsymbol{r}}\cdot d\boldsymbol{r} + \dfrac{\partial H}{\partial \boldsymbol{p}}\cdot d\boldsymbol{p} = 0 \quad (26)$$

leading to a Hamiltonian dynamical system of the form

$$\begin{cases} \dfrac{d\boldsymbol{r}}{dt} = \dfrac{\partial H}{\partial \boldsymbol{p}} = \dfrac{\boldsymbol{p}}{m} \\ \dfrac{d\boldsymbol{p}}{dt} = -\dfrac{\partial H}{\partial \boldsymbol{r}} = -\dfrac{\partial}{\partial \boldsymbol{r}}[V(\boldsymbol{r}) - \dfrac{\hbar^2}{2m}\dfrac{\nabla^2 R}{R}] \end{cases} \quad (27)$$

strictly similar to the ray-tracing system (13). If we envisage the system (27) for what it appears to be, without superimposing any interpretative prejudice, it is quite evident that its mathematical treatment is the same employed in the classical ray-tracing case, including the fact that the function *R(x,y,z)* is "transported" along the field lines of $\boldsymbol{p} \equiv \underline{\nabla}S$, to which $\underline{\nabla}R$ turns out to be perpendicular. The gradient $\dfrac{\partial}{\partial \boldsymbol{r}}\dfrac{\nabla^2 R}{R}$, in its turn, remains tangent to the wave-front, without acting on the *amplitude* of the particle velocity (but modifying, in general, its *direction*). The only possible amplitude changes could be due to the presence of an external potential *V(x,y,z)*.
Once more, thanks to its constancy along each trajectory, the function *R(x,y,z)* may be assigned on the launching surface from where the beam is assumed to start, and numerically built up step by step, together with its derivatives, in the whole region spanned by the motion of the beam.

**6. The unique dimensionless Hamiltonian system**

A quite expedient step is now the passage to the new, dimensionless variables $\underline{\xi}, \underline{\rho}, \tau$ defined as the ratio of *r*, *p* and *t*, respectively, with $\lambda_0 \equiv 2\pi\hbar/p_0$ for the space variables, $p_0 \equiv (2mE)^{1/2}$ for the momentum variables (so that $\rho_0 = 1$), and $\dfrac{\lambda_0}{p_0/m}$ for the time variable.

The equation system (27) takes up therefore the form

$$\begin{cases} \dfrac{d\underline{\xi}}{dt} = \underline{\rho} \\ \dfrac{d\underline{\rho}}{dt} = -\dfrac{\partial}{\partial \underline{\xi}}[\dfrac{V(\underline{\xi})}{2E} - \dfrac{1}{8\pi^2}G(\underline{\xi})] \end{cases} \quad (28)$$

with

$$G(\underline{\xi}) = \dfrac{1}{R}(\dfrac{\partial^2 R}{\partial \xi^2} + \dfrac{\partial^2 R}{\partial \eta^2} + \dfrac{\partial^2 R}{\partial \zeta^2});$$
$$\underline{\xi} \equiv (\xi,\eta,\zeta) \equiv (x/\lambda_0, y/\lambda_0, z/\lambda_0) \quad (29)$$

It may be observed that no direct reference is present, in the dimensionless form (28) assumed by the *quantum* dynamical system (27), to the mass of the moving particles, and not even to $\hbar$.

Let us also observe that *the same dimensionless form (28)* is taken up by the ray-tracing system (13) - relevant to the *classical* electromagnetic case - by simply assuming $\tau = \frac{c\,t}{\lambda_o}$ and replacing $\underline{\rho}$ with $\frac{\mathbf{k}}{k_o} \equiv \frac{\mathbf{v}_{ray}}{c}$ and $\frac{V(x,y,z)}{E}$ with $[1 - n^2(x,y,z)]$, in agreement with the relations (19).

Once assigned on the launching surface of the beam, the function $G(\underline{\xi})$ may be numerically determined step by step, in principle, together with its derivatives, by means of an interpolation process iterated along the full set of trajectories of the beam and connecting each step to the previous ones. This function, due to the wave amplitude distribution of the beam on the advancing wave-front, turns out to be the same - in correspondence with the same boundary conditions - for *classical* electromagnetic rays as well as for *quantum* material particles, although it has obviously nothing to do, in the electromagnetic case, with quantum features. In its absence, however, the system (28) would simply describe the *classical* motion of each particle of the beam. Due to the small coefficient $\frac{1}{8\pi^2}$, the transverse gradient $\frac{\partial G}{\partial \underline{\xi}}$ acts along the trajectory pattern in a soft and cumulative way: a fact granting the main justification for omitting such a term, as is done both in classical dynamics and in the standard geometrical optics approximation.

The trajectory pattern, in its turn, is a stationary structure determined at the very outset in a way somewhat reminding the spirit of classical variational principles, such as the ones of Fermat and Maupertuis. For any set of boundary conditions imposed to the function $R(x,y,z)$ on the launching surface of the beam, and for any assigned refractive medium (or force field), the system (28) provides both a "weft" of "rails" and a motion law to which particles (or rays) are deterministically bound, *showing no trace of probabilistic features*.

Each particle (as well as each electromagnetic ray) of the beam turns out to be conceivable, on the basis of the present analysis, as *starting and remaining* on a well definite trajectory. Such a trajectory belongs to a pattern which is *a priori* fixed, as a whole, by the properties of the medium and by the values assigned to the beam amplitude distribution $R(x,y,z)$ on the launching surface.

The system (28) provides, in conclusion, a set of dynamical laws which replace - and contain as a limiting case, when the transverse gradient $\frac{\partial}{\partial \underline{r}} \frac{\nabla^2 R}{R}$ may be assumed to be negligible - the classical ones.

In striking divergence from the classical dynamical laws, however, the new set of equations, because of its equivalence with a Helmholtz-like equation, requires in general the full set of boundary conditions for the determination of each trajectory of the beam.

## 7. Wave-like features in Hamiltonian form

Although an accurate and general numerical treatment lies beyond the aims of the present paper, we propose here the application of the equation system (28) to the propagation of a collimated beam injected at $\zeta = 0$ parallel to the $\zeta$-axis, and centred at $\xi = 0$, in order to simulate wave diffraction through a single slit.

The problem may be faced by taking into account for simplicity sake (but with no substantial loss of generality) either a (quantum) *particle beam* in the absence of external fields ($V = 0$), or a (classical) *electromagnetic beam* in vacuum ($n^2 = 1$), with a geometry allowing to limit the computation to the trajectories lying on the ($\xi\zeta$)-plane.

The Hamiltonian system (28) takes up therefore the form

$$\begin{cases} \dfrac{d\xi}{d\tau} = \rho_x \\ \dfrac{d\zeta}{d\tau} = \rho_z \\ \dfrac{d\rho_x}{d\tau} = \dfrac{1}{8\pi^2} \dfrac{\partial}{\partial \xi} G(\xi,\zeta) \\ \dfrac{d\rho_z}{d\tau} = \dfrac{1}{8\pi^2} \dfrac{\partial}{\partial \zeta} G(\xi,\zeta) \end{cases} \quad (30)$$

with

$$G(\xi,\zeta) = \frac{1}{R}\left(\frac{\partial^2 R}{\partial \xi^2} + \frac{\partial^2 R}{\partial \zeta^2}\right);$$

$$\rho_x(\zeta = 0) = 0;\ \rho_z(\zeta = 0) = \rho_0 = 1 \quad (31)$$

and a suitable amplitude distribution $R(\xi, \zeta=0)$ (from whose normalization the function $G$ is obviously independent) imposed at $\zeta = 0$.

Because of the transverse nature of the gradient of $G(\xi,\zeta)$, the *amplitude* of the vector $\underline{\rho}$ remains unchanged (in the absence of external fields and/or refractive effects) along each trajectory, leading therefore to the relation

$$\rho_z = \sqrt{\rho_o^2 - \rho_x^2} \equiv \sqrt{1 - \rho_x^2}, \quad (32)$$

which may advantageously replace the fourth equation of the Hamiltonian system (30). Two possible models of the amplitude distribution $R(\xi, \zeta=0)$ are obtained by assuming either

• a Gaussian distribution centred at $\xi = 0$, in the form

$$R_0(\xi; \zeta = 0) \div e^{-\left(\frac{x}{w_0}\right)^2} \equiv e^{-\varepsilon^2 \xi^2} \quad (33)$$

(with constant $w_0$ and $\varepsilon = \frac{\lambda_0}{w_0} \leq 1$), a functional form suggested by its smooth analytical behaviour; or

• algebraic distributions, in the form

$$R_N(\xi; \zeta = 0) \div \frac{1}{1 + \left(\frac{x}{w_0}\right)^{2N}} \equiv \frac{1}{1 + (\varepsilon\,\xi)^{2N}} \quad (34)$$

(with integer *N*), allowing the presence of flat central regions, widening with increasing *N*. We show in **Fig.1** for $\varepsilon = 0.25$, both the (Gaussian) distribution $R_0$ and the (algebraic) distributions $R_{1,2}$ (with *N=1* and *N=2*, respectively).

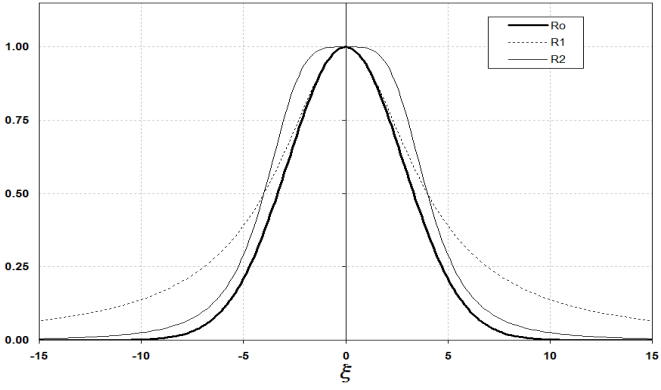

**Fig.1** - Plot of the amplitude distributions $R_{0,1,2}$ assigned to the beam on the launching plane $\zeta = 0$, for $\varepsilon = \frac{\lambda_0}{w_0} = 0.25$,

- in the Gaussian case of eq.(33) (continuous heavy line) and
- in the algebraic cases of eq.(34), with $N=1$ (dotted line) and $N=2$ (continuous light line), respectively.

**Fig.2** and **Fig.3** represent, in their turn, the functions

$$G_{1,2}(\xi;\zeta=0) = \frac{1}{R_{1,2}} \frac{d^2 R_{1,2}}{d\xi^2} \quad (35a)$$

respectively, each one compared with

$$G_0(\xi;\zeta=0) = \frac{1}{R_0} \frac{d^2 R_{1,2}}{d\xi^2} \quad . \quad (35b)$$

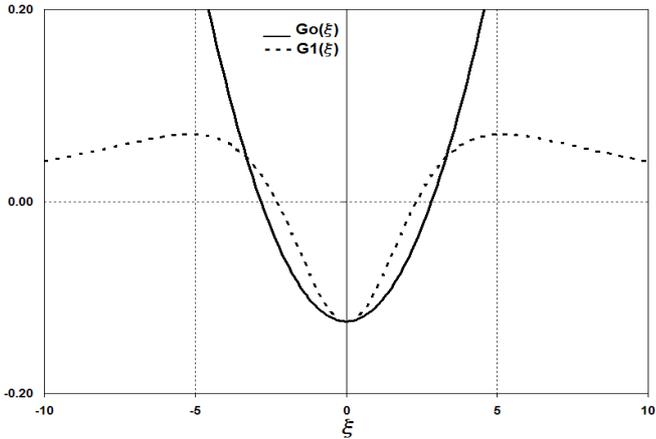

**Fig.2** - Plot of the initial functions $G_{0,1}$ of eqs. (35 a,b) corresponding to the distributions $R_{0,1}$ of **Fig.1**, respectively.

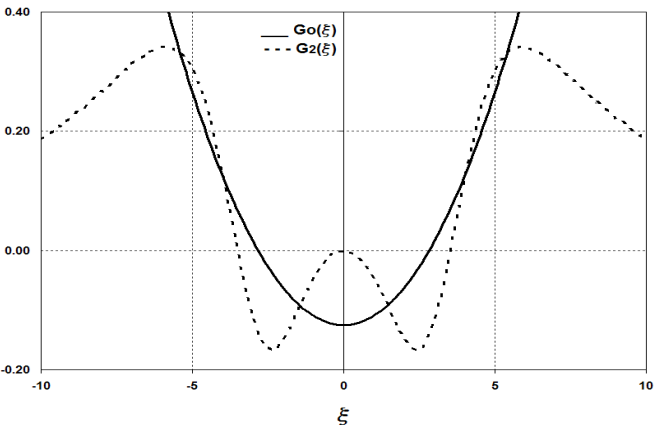

**Fig.3** - Plot of the initial functions $G_{0,2}$ of eqs. (35 a,b) corresponding to the distributions $R_{0,2}$ of **Fig.1**, respectively.

**Figs.4-6** present, finally, the beam trajectories starting with $R_{0,1,2}(\xi;\zeta=0)$, respectively.

The functions $G_{0,1,2}(\xi;\zeta>0)$ were built up step by step by means of a 3-points Lagrange interpolation. As predicted by the standard optical diffraction theory [14], no "fringe" is found in the Gaussian case of **Fig.4** (due to the fact that the Fourier transform of the distribution $R_o$ consists of another Gaussian function), while "fringes" appear (in the form of gathering trajectories) in **Figs.5** and **6** for the algebraic initial distribution $R_{1,2}$, focussing closer to the launching plane for higher values of $\varepsilon$.

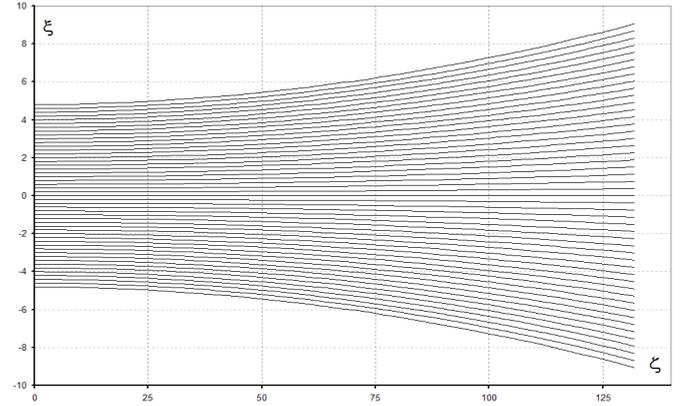

**Fig.4** - Trajectory pattern on the $(\xi,\zeta)$-plane, in the Gaussian case of **Fig.1**.

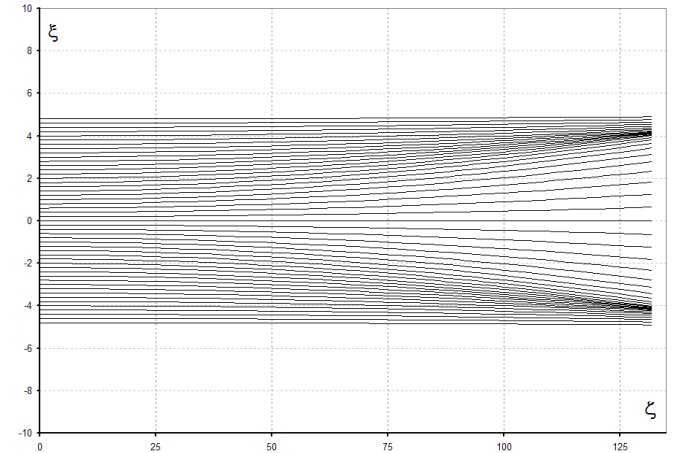

**Fig.5** - Trajectory pattern on the $(\xi,\zeta)$-plane, in the algebraic case $N=1$ of **Fig.1**.

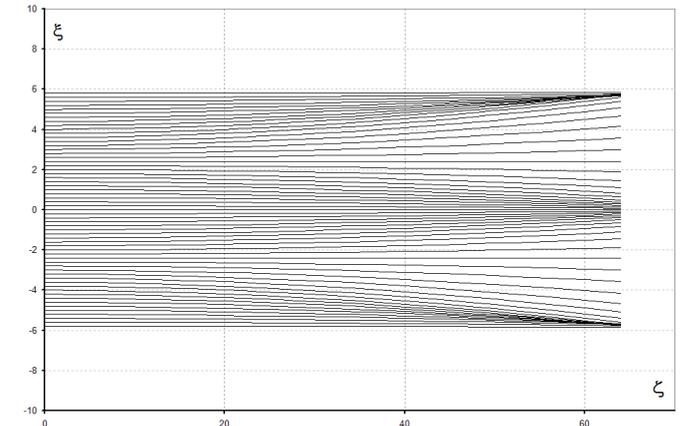

**Fig.6** - Trajectory pattern on the $(\xi,\zeta)$-plane, in the algebraic case $N=2$ of **Fig.1**.

While the basic result to be pointed out here is the very appearance of fringes in the context of our Hamiltonian approach, we stress once more their strong dependence on the beam launching conditions.

No further difficulty would be encountered in the case of *two* beams, injected parallel to the $\zeta$-axis at $\zeta = 0$ and centred, on the $\xi$-axis, at two symmetrical points $\xi = \pm \xi_0$, in order to simulate both their diffraction and their interference through a double slit.

## 8. Discussion and conclusions

A certain analogy may be observed between the results of the present work and the ones previously published by one of the Authors (A.O.) in a quite different context [15-17]. Another obvious analogy is found with Refs.[6, 7] (based on **Bohm**'s approach) which are hindered, however, by a highly non-linear Hamilton-Jacobi set of equations needing, in general, an often unattainable *generating function*: an obstacle which is avoided by a solution method requiring the previous knowledge of the wave function $\psi$. Such an entangled procedure should be compared with our directly integrable set of Hamiltonian motion laws.

Let us stress, incidentally, that we analyze here, for the first time, the *diffractive case*, and show the *strong dependence* both of the ray geometry and of the motion law on the transverse intensity profile of the beam (our eqs.(33) and (34)).

The "fixed weft of rails" (i.e. the trajectory pattern), on the other hand, does not depend (for not-interacting particles or rays), on the intensity of the beam, which could even be reduced to a single particle at a time: a peculiar feature which has often (and somewhat misleadingly) induced to speak of single particle *self-diffraction*.

We would like to conclude the present paper by tackling one of its conceivable developments. Let us remind that one of the possible ways to support quantum non-locality is the assertion that if, in a two-slits quantum interference experiment, one of the slits is removed, a single diffraction pattern is replaced to the previous one, reaching each particle previously passed through the slits, and affecting its motion in an immediate and non-local way.

Imagine now, for instance, to keep one of the slits fixed, while moving away the other one, and recall that, as we have shown, the "wefts" of "rails" are exactly the same both in classical and in quantum experiments.

In any case the initial pattern shall be progressively changed into a single-slit diffraction one; but the "non-local" information which we may associate, at any time, to any point of space is only the virtual geometrical form of the pattern due to the instantaneous position of the moving slit: a merely mathematical knowledge which, in itself, has no physical relevance at all, and requires no transport of energy or information, since it is present *only in our mind*. Any physical information, in fact, appears to be carried along by the rays and/or particles, with their characteristic velocities: this would be stated in the **classical** case, and the same statement should be logically made in the **quantum** case, sharing with the previous one the same mathematical and logical frame.